\documentclass{webofc}

\usepackage[varg]{txfonts}   
\usepackage{hyperref}
\usepackage{url}
\usepackage{comment}
\usepackage{slashed}
\hypersetup{colorlinks=true,citecolor=blue,urlcolor=blue,linkcolor=blue}
\begin{document}
\title{Top quark flavor changing neutral currents at Future Linear Colliders}
%
%

\author{\firstname{Adil} \lastname{Jueid}\inst{1}\fnsep\thanks{Speaker,~\email{adiljueid@ibs.re.kr}} \and
        \firstname{Shinya} \lastname{Kanemura}\inst{2}\fnsep\thanks{\email{kanemu@het.phys.sci.osaka-u.ac.jp}}
}

\institute{Particle Theory and Cosmology Group, Center for Theoretical Physics of the Universe, Institute for Basic Science (IBS),
Daejeon, 34126, Republic of Korea 
\and
           Department of Physics, Osaka University, 
Toyonaka, Osaka 560-0043, Japan 
}

\abstract{We discuss the production and the decay of top quark through flavor-changing neutral current (FCNC) interaction at future linear colliders. We first discuss the theoretical predictions of top quark FCNC decays into $qH$ and $qZ$ within a class of $t$--channel simplified dark matter models. For the existing bounds on the top quark FCNC interactions at the Large Hadron Collider, we estimate the production rates of top quark through FCNC interactions at future linear colliders for energies from $250$ GeV to $3$ TeV.}
\maketitle
%
\section{Introduction}
\label{sec:intro}

In the Standard Model (SM) of particle physics, the flavor changing neutral current (FCNC) interactions are suppressed as a consequence of the Glashow-Iliopoulos-Maiani (GIM) mechanism \cite{Glashow:1970gm}. Hence, the rates of {\it e.g.} top quark FCNC decays are extremely small in the SM \cite{Eilam:1990zc}. It is found that the branching ratios of top quark FCNC decays into $qX$ ($X=\gamma,Z,g,H$) range from about $10^{-17}$ to $10^{-12}$ making the discovery of this phenomena at the LHC impossible if one assumes the SM. Several extensions of the SM predict quite sizeable rates of top quark FCNC decays \cite{Hou:1991un,Atwood:1996vj,Lopez:1997xv,Yang:1997dk,Eilam:2001dh,Aguilar-Saavedra:2004mfd,Gaitan:2004by,Frank:2005vd,Bejar:2006ww,Agashe:2006wa,Cao:2007dk,Baum:2008qm,Han:2009zm,Agashe:2009di,Gao:2013fxa,Dedes:2014asa,Abbas:2015cua,Botella:2015hoa,Dey:2016cve,Diaz-Furlong:2016ril,Hung:2017tts,Kim:2018oih,Banerjee:2018fsx,Bolanos:2019dso,Liu:2021crr,Chen:2022dzc,Crivellin:2022fdf,Chen:2023eof,Frank:2023fkc,Jueid:2024cge}. The relatively large top quark FCNC rates compared to the SM predictions have motivated the ATLAS and CMS collaborations to carry several searches targeting FCNC phenomena in both the production ($pp \to t X$) and the decay ($p p \to t\bar{t} \to q X + b W$) stages -- see for example Refs. \cite{CMS:2021gfa,ATLAS:2022per,ATLAS:2023qzr,ATLAS:2023ujo} for the most recent results --. Unfortunately, no signal beyond the SM backgrounds was observed and relatively strong bounds on the different top quark FCNC branching ratios were imposed\footnote{We also note that top quark FCNC phenomena has motivated various theoretical studies for future hadron and lepton colliders (see {\it e.g.} Refs. \cite{Khatibi:2021phr,Liu:2019wmi,Bhattacharya:2023beo,Liu:2020kxt,Oyulmaz:2019jqr}).}.

Top quark FCNC phenomena may be connected to the nature of neutrino mass, the mechanism of electroweak symmetry breaking or the dark matter (DM) in the universe. We first consider the possibility that DM generates sizeable rates for top quark FCNC decays at the one-loop order. As a benchmark model, we study the case of $t$--channel simplified DM model that extends the SM with two $Z_2$ odd particles: a scalar mediator ($S$) carrying the same quantum numbers as of a right-handed up-type quark and a right-handed fermion ($\chi$). In this setup, the mediator $S$ is a colored scalar and couples to all the quark generations which therefore implies nonzero contributions to top quark FCNC decays at the one-loop order (see section \ref{sec:topFCNC})\footnote{The phenomenology of this class of models has been extensively studied in the literature (see e.g. Refs. \cite{Mohan:2019zrk,Arina:2020udz,Arina:2020tuw,Becker:2022iso,Arina:2023msd}). In these studies, it was assumed that each generation of quarks couple to one mediator only in order to avoid the flavor constraints.}. We study in particular the decays $t\to qZ$ and $t\to qH$ -- where $q=u,c$ -- in detail and we found that the corresponding branching ratios are related to each other and independent of the light quark mass. 

We also discuss the production cross sections for top quark FCNC phenomena at future linear colliders for center-of-mass energies from $250$ GeV to $3$ TeV. The analysis is done in a model-independent way and using values of the couplings that are allowed by the current bounds. We study processes where top quark FCNC phenomena is involved in the production stage ($e^+ e^- \to q \bar{t} + {\rm h.c.}$) and in the decay stage ($e^+ e^- \to t \bar{t} \to q X b W$ where $X=\gamma,Z,g, H$). 

The remainder of this article is as follows. In section \ref{sec:topFCNC} we analyse the top quark FCNC decays generated by the loops of DM sector particles. We discuss the production rates for processes involving top quark FCNC interactions at future linear colliders in section \ref{sec:topFCNC:LCs}. We draw our conclusions in section \ref{sec:conclusions}.

\section{Top quark FCNC decays triggered by dark matter}
\label{sec:topFCNC}

We consider a $t$--channel simplified model which extends the SM with a colored scalar mediator ($S$) and a right-handed fermion ($\chi$) both are odd under an ad-hoc $Z_2$ symmetry \cite{Jueid:2024cge}. Under $SU(3)_c \otimes SU(2)_L \otimes U(1)_Y$, the new particles transform as 
\begin{eqnarray}
S: ~ ({\bf 3}, {\bf 1})_{+2/3}, \quad \chi: ~ ({\bf 1}, {\bf 1})_{0}.
\end{eqnarray}
In this framework, the DM particle interacts primarily with right-handed up-type quarks through a Yukawa-type interaction. Under these symmetry requirements, the most general Lagrangian is given by
\begin{eqnarray}
{\cal L} \supset {\cal L}_{\rm S} + {\cal L}_{\chi} - V(S, \Phi),
\end{eqnarray}
with ${\cal L}_{S}$, ${\cal L}_{\chi}$ and $V(S, \Phi)$ being the kinetic Lagrangian of the mediator, the Yukawa-type Lagrangian of the DM particle and the scalar potential respectively. They are given by
\begin{eqnarray}
{\cal L}_S + {\cal L}_\chi \equiv i \bar{\chi} \slashed{\partial} \chi^c + \frac{1}{2} M_\chi \bar{\chi} \chi^c + ({\cal D}_\mu S)^\dagger ({\cal D}^\mu S) + \sum_{q=u,c,t} \bigg(Y_q \bar{q}^c_R \chi S + {\rm h.c.} \bigg).
\label{eq:Lag:S}
\end{eqnarray}
Here, the first and the second terms refer to the kinetic and mass terms of the right-handed fermion, the third term to the kinetic term of $S$ and the last term to the Yukawa-type interaction of the dark sector particle to the right-handed up-type quarks. The most general, {\it CP} conserving, renormalizable and gauge-invariant scalar potential is given by 
\begin{eqnarray}
V(S, \Phi) = - m_{11}^2 |\Phi|^2 + m_{22}^2 |S|^2 + \lambda_1 |\Phi|^4 + \lambda_2 |S|^4 + \lambda_3 |S|^2 |\Phi|^2,
\end{eqnarray}
with $\Phi$ being the SM Higgs doublet. Note that $\lambda_2$ does not affect the phenomenology of the model and therefore will be set to one without loss of generality.

\begin{figure}[!t]
\centering
\includegraphics[width=0.8\linewidth]{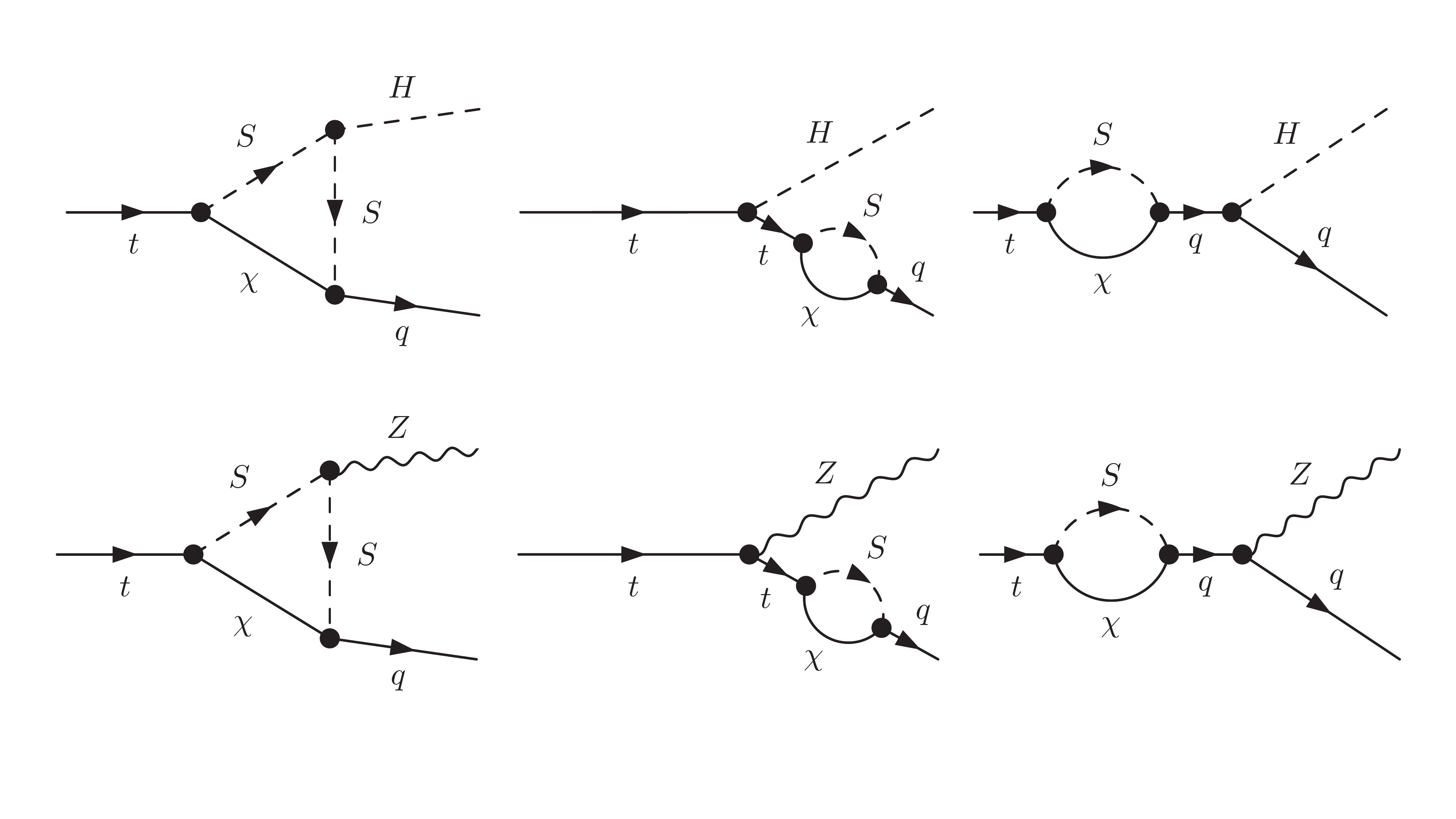}
\vspace{-1cm}
\caption{Examples of Feynman diagrams for the $t\to q H$ and $t\to q Z$ decays at the one-loop order.}
\label{fig:FCNC:decay}
\end{figure}

We consider the two-body decays of the top quark into $qZ$ and $qH$ ($q=u,c$). Examples of Feynman diagrams at the one-loop order are shown in Fig. \ref{fig:FCNC:decay}. The effective operator for these top quark FCNC decays can be generically paramterised as 
\begin{eqnarray}
- {\cal L}_{\rm eff} &=& \bar{t} \gamma^\mu (f_{tqZ}^L P_L + f_{tqZ}^R P_R) q Z_\mu +  \bar{t} p^\mu (g_{tqZ}^L P_L + g_{tqZ}^R P_R) q Z_\mu \nonumber \\
&+& \bar{t} (f_{tqH}^L P_L + f_{tqH}^R P_R) q H + {\rm h.c.},
\end{eqnarray}
with $f_{tqX}^{L,R}~(X=Z,H)$ and $g_{tqZ}^{L,R}$ being the form factors, $P_{L,R}=(1\mp \gamma_5)/2$ are the projection operators and $p^\mu$ is the 4-momentum of the top quark. In the limit of vanishing light quark masses, we have
\begin{eqnarray}
f_{tqH}^L &\approx&  \frac{3 Y_q Y_t m_t}{16 \pi^2} \lambda_3 \upsilon C_1, \qquad f_{tqH}^R \approx 0, \nonumber \\
f_{tqZ}^L &\approx& 0, \qquad f_{tqZ}^R \approx -\frac{g_1 s_W Y_q Y_t}{24 \pi^2} \bigg(2 C_{00} + B_{1,t}\bigg), \\ 
g_{tqZ}^L &\approx& \frac{g_1 s_W Y_q Y_t m_t}{12 \pi^2} \bigg(C_1 + C_{11} + C_{12}  \bigg), \qquad
g_{tqZ}^R \approx 0, \nonumber
\label{eq:tqH:equation}
\end{eqnarray}
where $B_{1,t} \equiv B_{1}(m_t^2, M_\chi^2, M_S^2)$ and $C_{i,ij} \equiv C_{i,ij}(m_t^2, M_X^2, m_q^2, M_\chi^2, M_S^2, M_S^2)$ refer to the 2- and 3-point scalar Passarino-Veltman functions. The expressions of the form factor have been obtained using \textsc{FeynArts} version 3.11 \cite{Hahn:2000kx} and \textsc{FormCalc} version 9.9 \cite{Hahn:2006zy}. We furthermore use \textsc{LoopTools} version 2.16 for their numerical evaluation \cite{Hahn:1999mt}. The resulting branching ratios are given by 
\begin{eqnarray}
{\rm BR}(t \to q X) = \frac{\Gamma(t \to qX)}{\Gamma_t},
\end{eqnarray}
with $\Gamma_t \equiv \Gamma(t\to bW) = 1.32$ GeV being the total width calculated at Next-to-next-to-LO in QCD \cite{Gao:2012ja} and $\Gamma(t\to qX)$ is the corresponding top FCNC partial width \cite{Jueid:2024cge}. Note that our model predicts the following relations
\begin{eqnarray}
\frac{\Gamma(t\to c X)}{\Gamma(t\to u X)} \approx \left(\frac{Y_c}{Y_u}\right)^2, \qquad     \frac{\Gamma(t \to q Z)}{\Gamma(t \to q H)} \approx \frac{{\cal O}(10)}{\lambda_3^2}.
\end{eqnarray}

\begin{figure}[!t]
\centering
\includegraphics[width=0.48\linewidth]{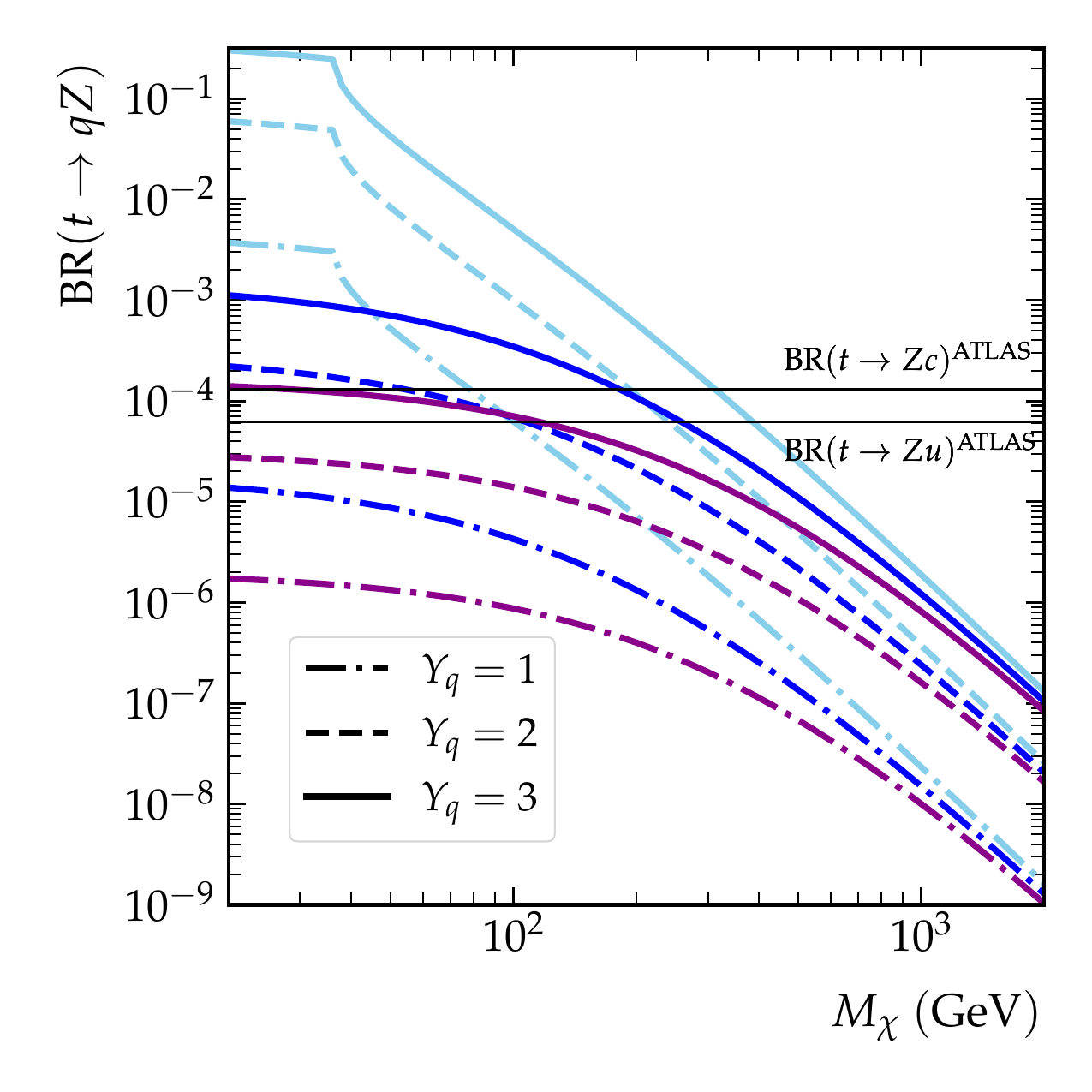}
\hfill
\includegraphics[width=0.48\linewidth]{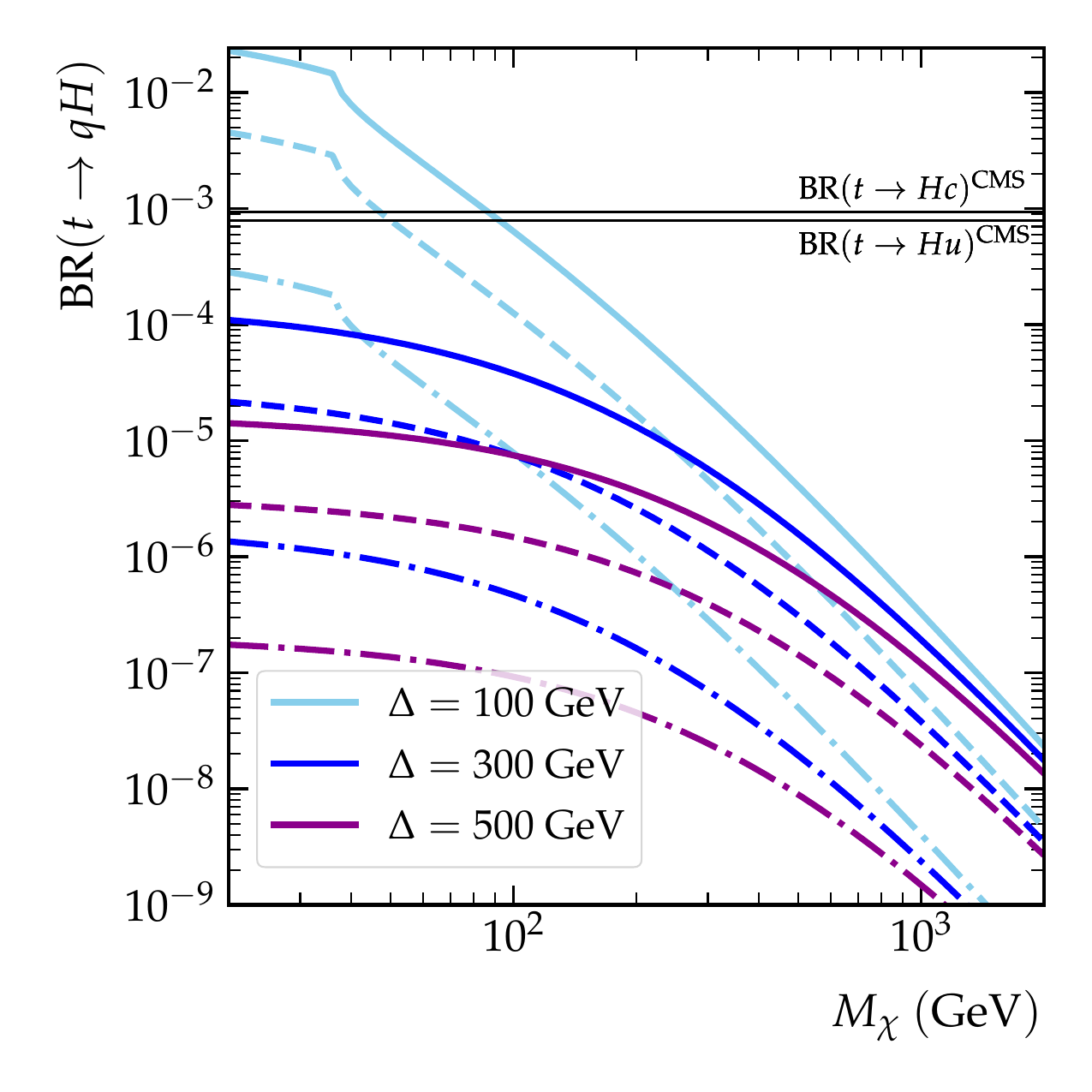}
\caption{The FCNC decay branching ratios as function of $M_\chi$ for $t \to q Z$ (left) and $t \to q H$ (right). The  results are shown for $Y_q = 1, 2$ and $3$ and $\Delta \equiv M_S - M_\chi = 100, 300$ and $500$ GeV. Here, we assume that $Y_q \equiv Y_u = Y_c = Y_t$ and $\lambda_3 = 1$. We also show in the same panels the latest exclusion bounds from the searches of FCNC phenomena reported on by the ATLAS \cite{ATLAS:2023qzr} and the CMS \cite{CMS:2021gfa} collaborations.}
\label{fig:BRtqX}
\end{figure}

In Fig.  \ref{fig:BRtqX}, we show the branching ratios of $t\to q Z$ (left) and of $t \to q H$ (right) as a function of $M_\chi$ for different values of the mass splitting $\Delta$ and of the Yukawa-type coupling $Y_u=Y_c=Y_t$. We can see that, since these two branching ratios scale as $|Y_q Y_t|^2$, higher values of $Y_q = Y_t$ will lead to extremely large branching ratios especially for small $M_\chi$ and small $\Delta$ (giving the $1/M^3$ dependence). Light DM masses of order $\approx 100$ GeV are still allowed if one considers $Y_q \approx 1$ and relatively large $\Delta$. Finally, the branching ratios do not depend on $\Delta$ for heavy DM ($> 1000$ GeV or so). Note that for phenomenologically viable benchmark scenarios (see Ref. \cite{Jueid:2024cge}) the maximum allowed values of the top FCNC branching ratios are of order $10^{-6}$--$10^{-9}$. 

\begin{figure}[!h]
    \centering
    \includegraphics[width=0.8\linewidth]{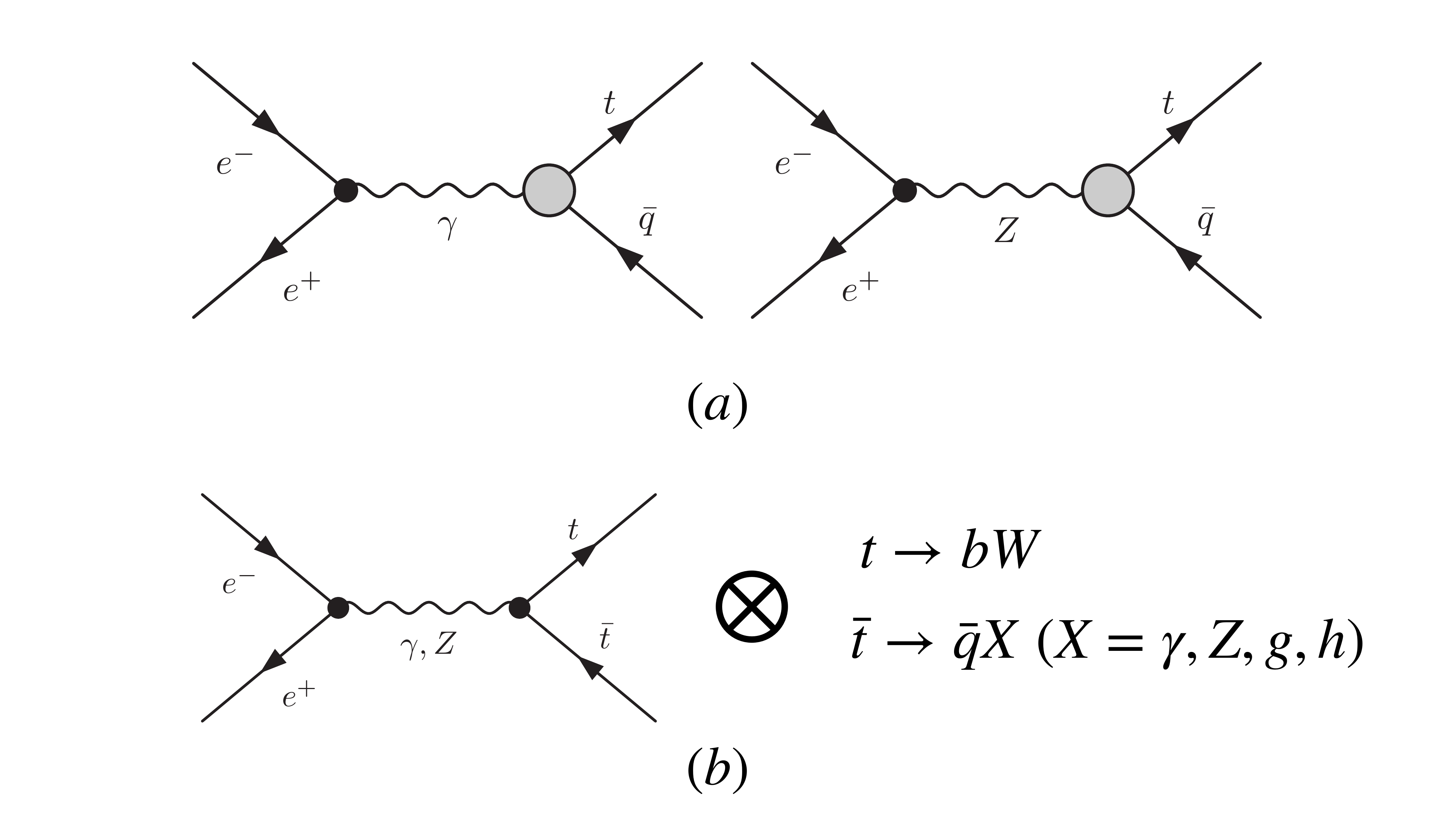}
    \caption{Examples of Feynman diagrams for the top quark FCNC phenomena in $e^+ e^-$ collisions. Here we show the FCNC processes at the production stage (upper panels) and at the decay state (lower panels).}
    \label{fig:FD:FCNC:production}
\end{figure}

\section{Top quark FCNC at future linear colliders} 
\label{sec:topFCNC:LCs}

We now turn into a brief discussion of the top quark FCNC rates at future linear colliders. We categorise the relevant processes into 
\begin{itemize}
    \item FCNC phenomena in the production stage in which case the top FCNC interaction vertices are involved at the production level.
    \item FCNC phenomena in the decay stage in which case top quark pairs are produced through electroweak interactions and then one (anti)top decays into $bW$ and the other (anti)top decays into $qX$ ($X=\gamma,Z,H,g$).
\end{itemize}
Examples of Feynman diagrams are shown in Fig. \ref{fig:FD:FCNC:production}. To estimate the production rates, we consider the most generic dimension five Lagrangian which is based on slightly modified implementation of Refs. \cite{Aguilar-Saavedra:2008nuh,Aguilar-Saavedra:2009ygx}\footnote{Other possible parameterisation can be found in Ref. \cite{Durieux:2014xla}.}
\begin{eqnarray}
    {\cal L} &=& \frac{g_s}{\Lambda} \bar{q} \lambda^a \sigma^{\mu\nu} \bigg(\zeta_{qt}^L P_L + \zeta_{qt}^R P_R\bigg) t G_{\mu\nu}^a - \frac{1}{\sqrt{2}} \bar{q}\bigg(\eta_{qt}^L P_L + \eta_{qt}^R P_R \bigg) \nonumber \\
    &-& \frac{g_W}{2 c_W} \bar{q} \gamma^\mu \bigg(X_{qt}^L P_L + X_{qt}^R P_R\bigg) t Z_\mu + \frac{g_W}{2 c_W \Lambda} \bar{q} \sigma^{\mu\nu} \bigg(K_{qt}^L P_L + K_{qt}^R P_R) t Z_{\mu\nu} \\
    &+& \frac{e}{\Lambda} \bar{q} \sigma^{\mu\nu} \bigg(\lambda_{qt}^L P_L + \lambda_{qt}^R P_R) t A_{\mu\nu}, \nonumber
    \label{eq:FCNC:general}
\end{eqnarray}
where $\Lambda$ is a new physics scale which we will set equal to $1$ TeV.  The current bounds on the top quark FCNC branching ratios from LHC searches can be translated into bounds on the effective couplings that appear in eq. \ref{eq:FCNC:general}. For the $tqZ$ coupling, the situation is a bit tricky as there are four independent form factors. In what follows we consider two scenarios: ({\it i}) the $tqZ$ vertex is purely vectorial ($K_{qt}^L = K_{qt}^R=0$) and ({\it ii}) the $tqZ$ vertex is purely tensorial ($X_{qt}^L = X_{qt}^R = 0$). For the case of top quark FCNC phenomena that occur in the decay stage, the total rate can be generically defined as
\begin{eqnarray}
    \sigma &\equiv&  2 \times \sigma(e^+ e^- \to t\bar{t}) \times {\rm BR}(t\to bW) \times {\rm BR}(t\to q X), \nonumber \\ 
    &\approx& 2 \times \sigma(e^+ e^- \to t\bar{t}) \times {\rm BR}(t\to qX),
\end{eqnarray}
since ${\rm BR}(t\to bW) \approx 100\%$. 

\begin{figure}[!t]
    \centering
    \includegraphics[width=0.49\linewidth]{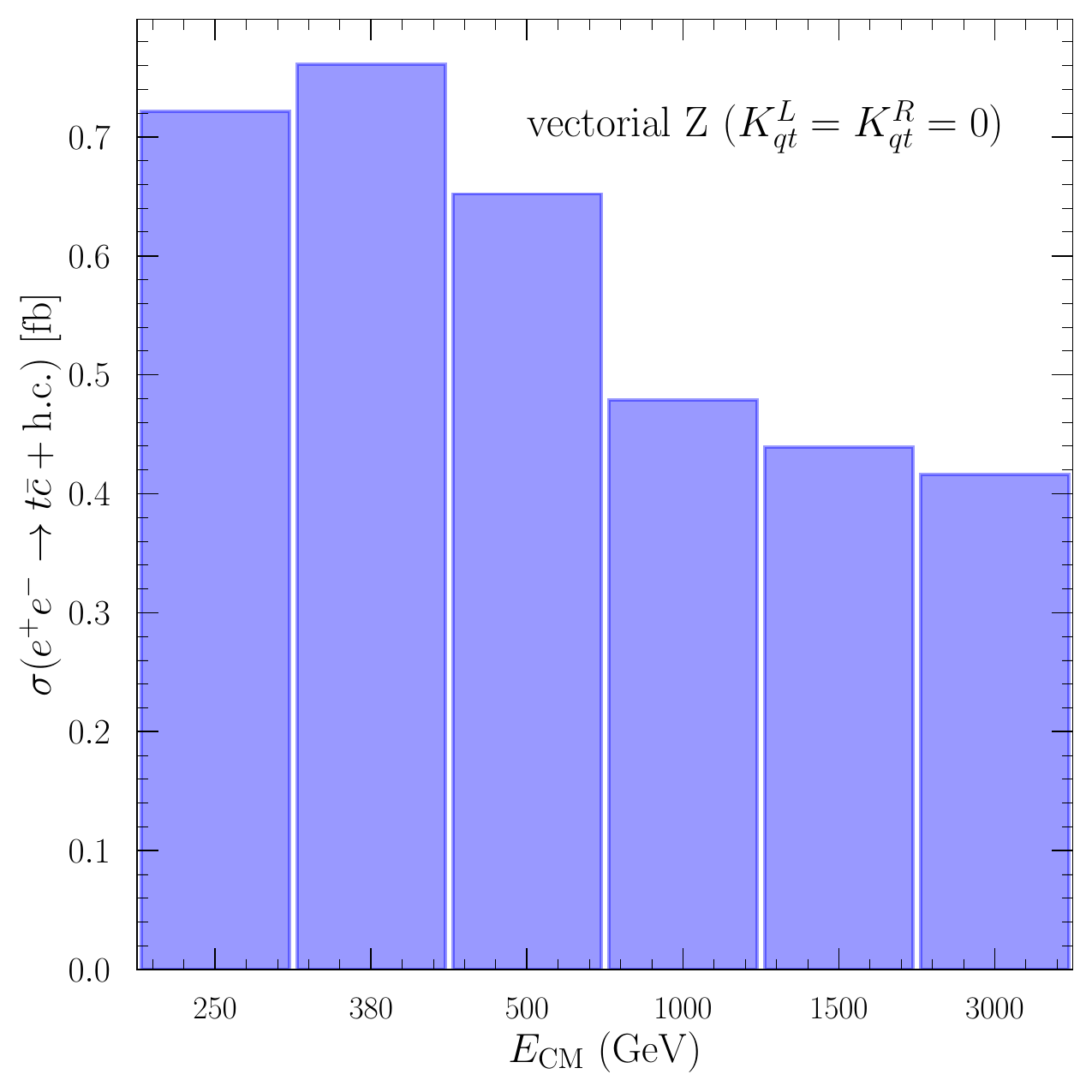}
    \includegraphics[width=0.49\linewidth]{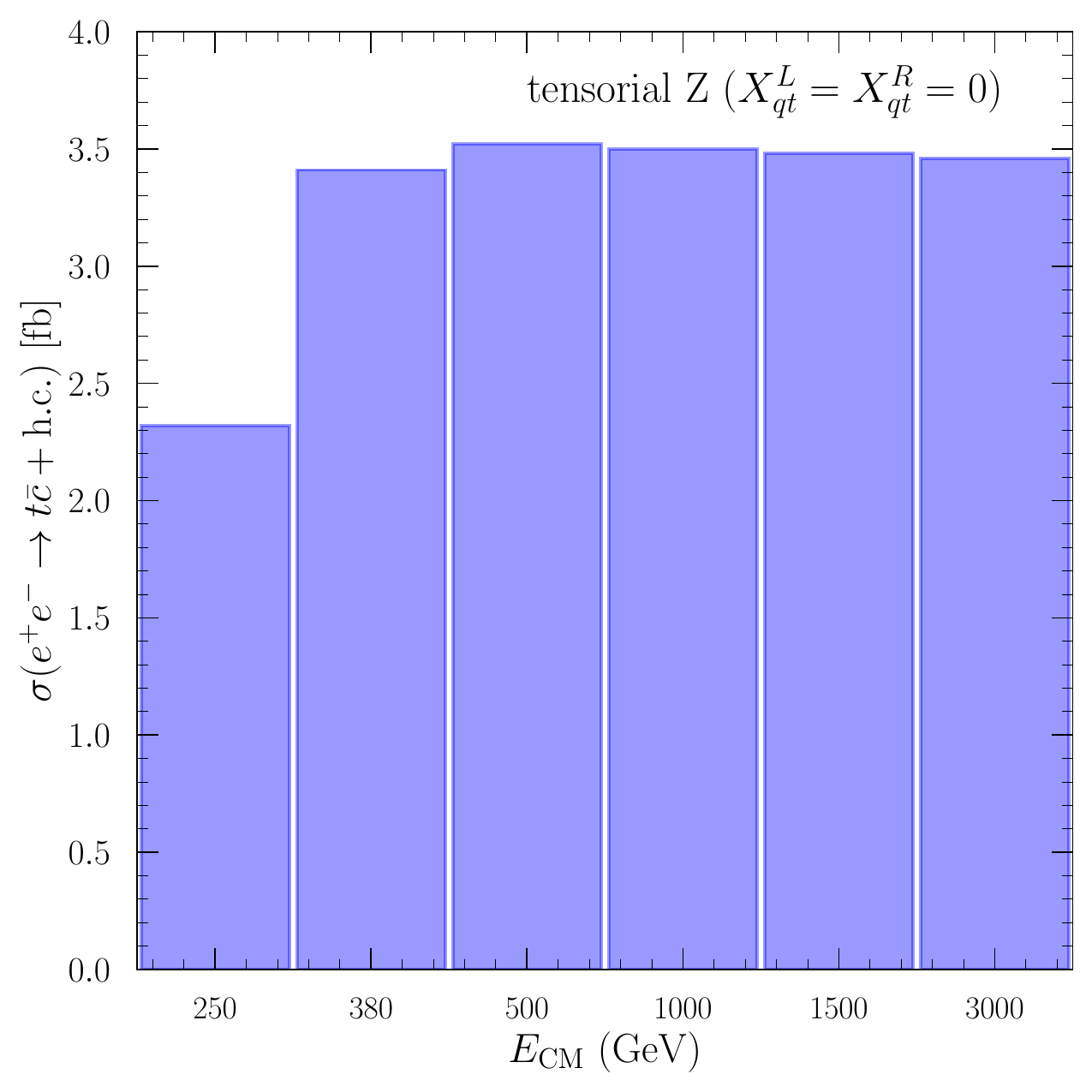}
    \caption{The total cross section for the production of a top (anti)quark in association with a charm quark in $e^+ e^-$ as a function of the center-of-mass energy. Here we show the results in the case where the $tqZ$ coupling is purely vectorial (left) and is purely tensorial (right). All the calculations have been done at the lowest order in perturbation theory.}
    \label{fig:sigma:ee:tc}
\end{figure}

\begin{figure}[!t]
    \centering
    \includegraphics[width=0.49\linewidth]{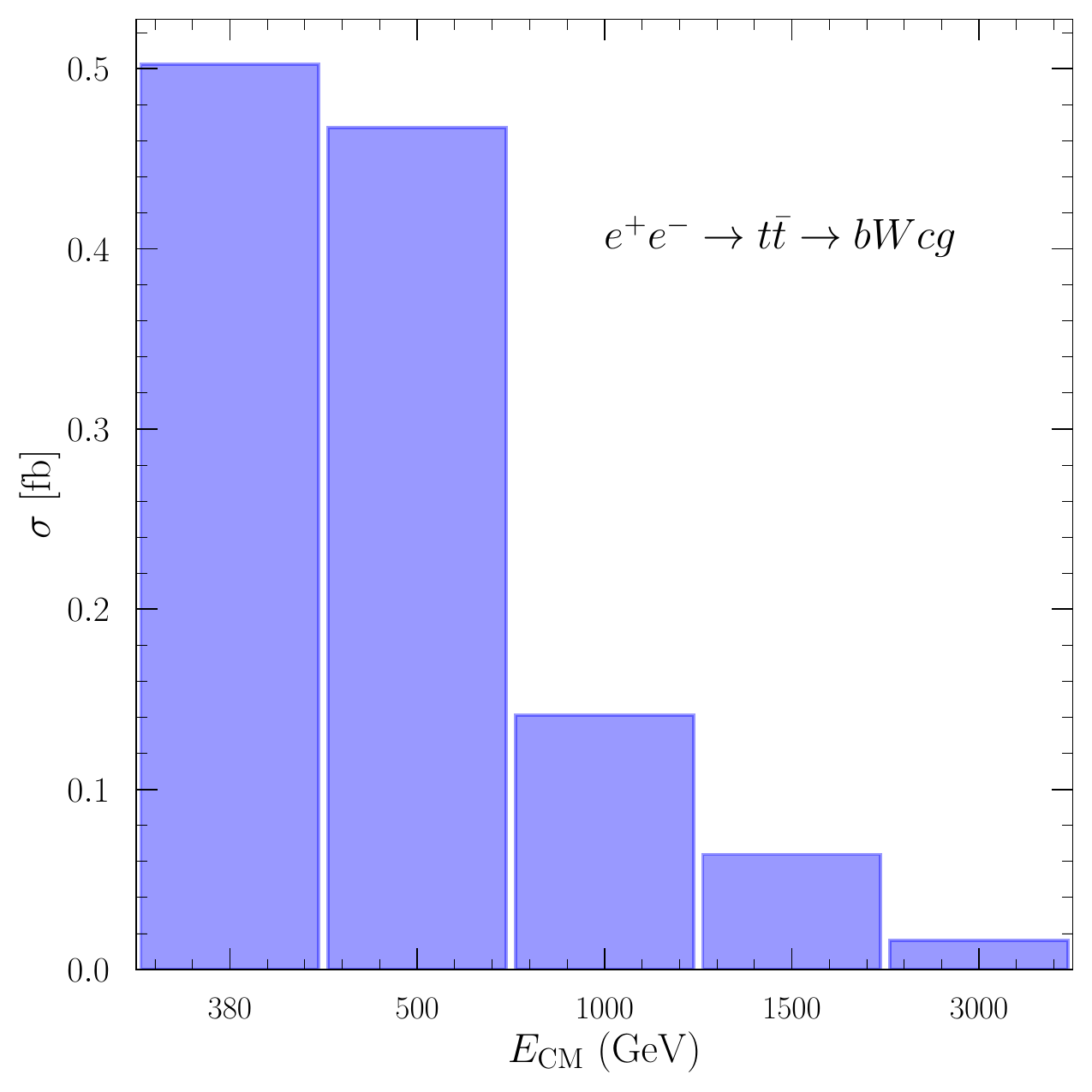}
    \includegraphics[width=0.49\linewidth]{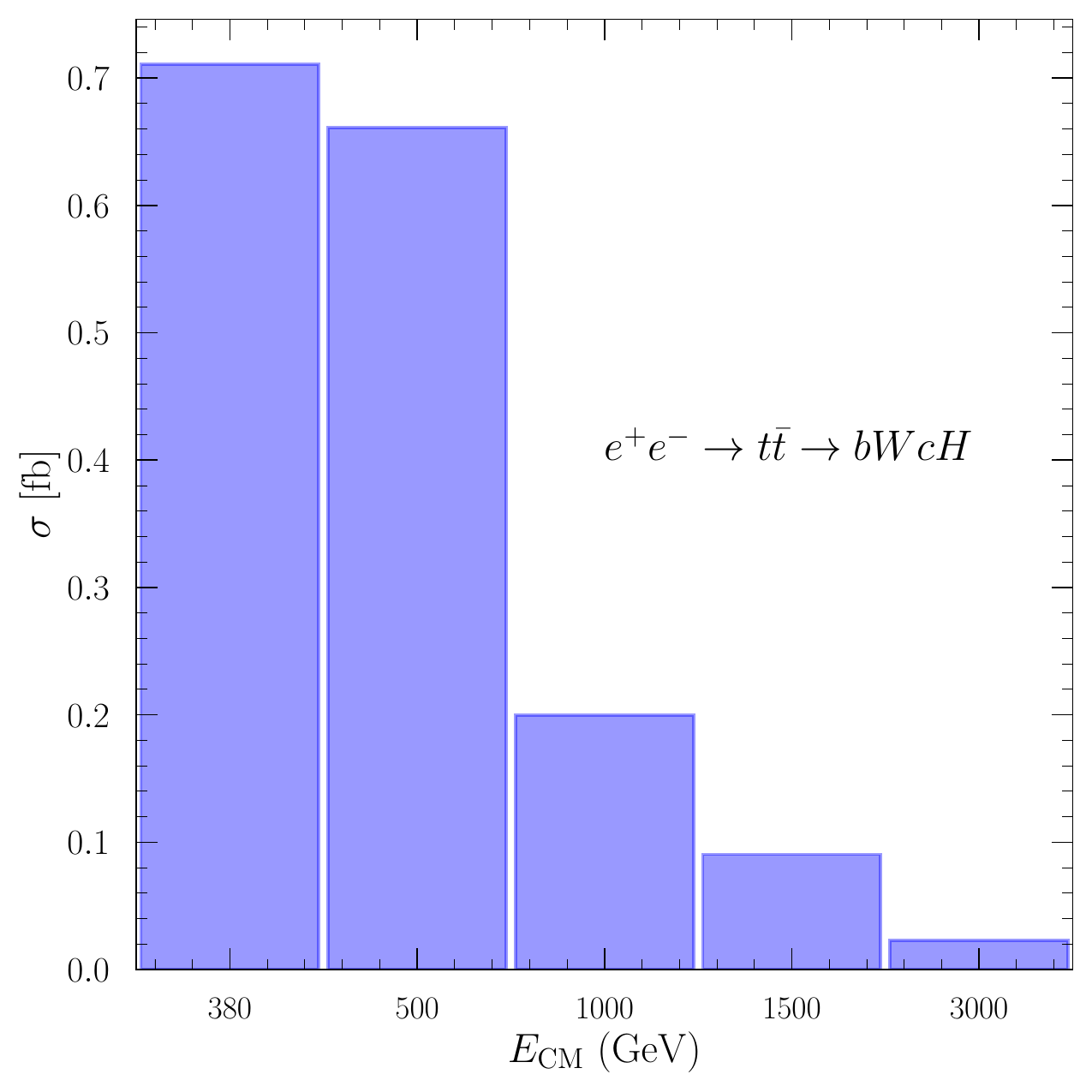}
    \includegraphics[width=0.49\linewidth]{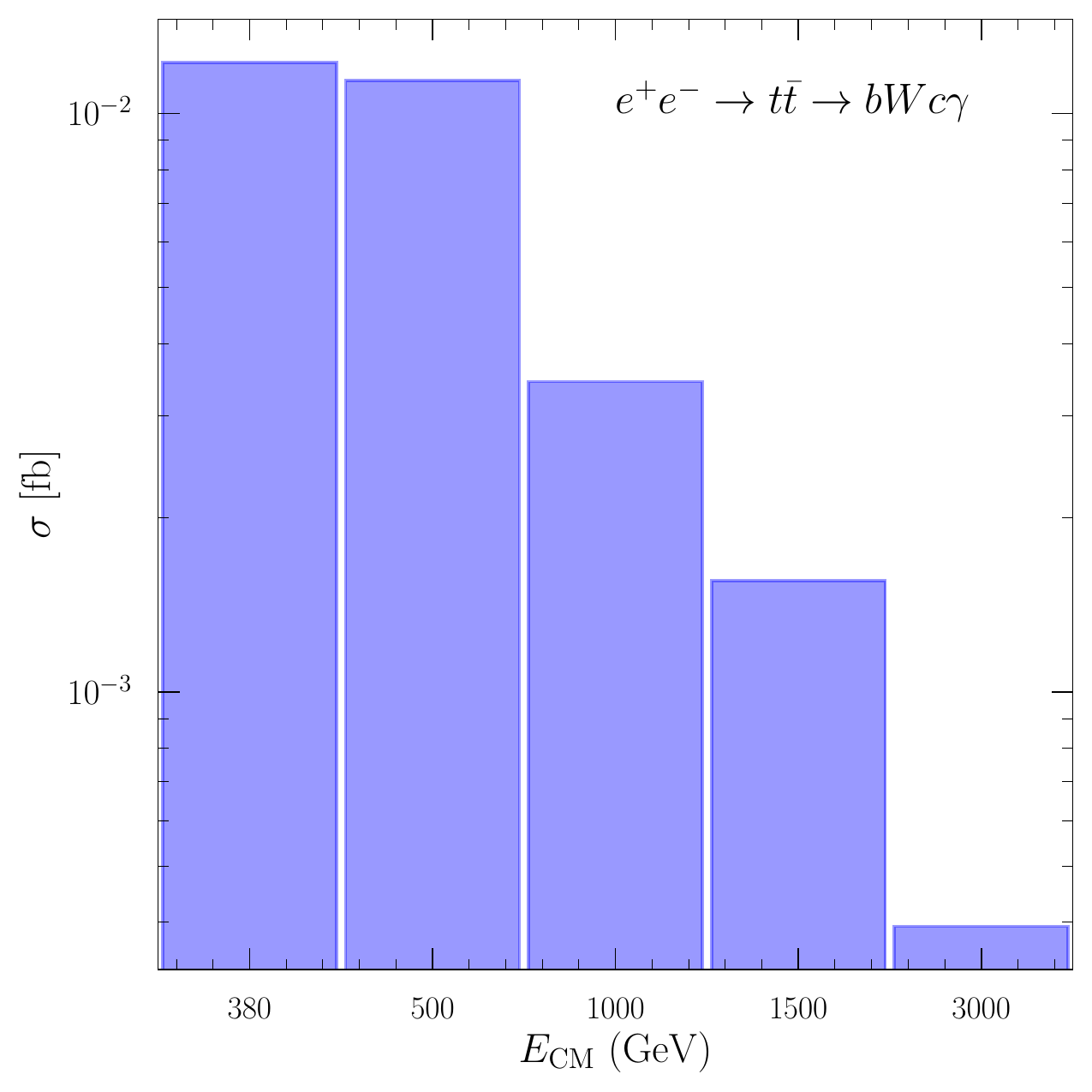}
    \includegraphics[width=0.49\linewidth]{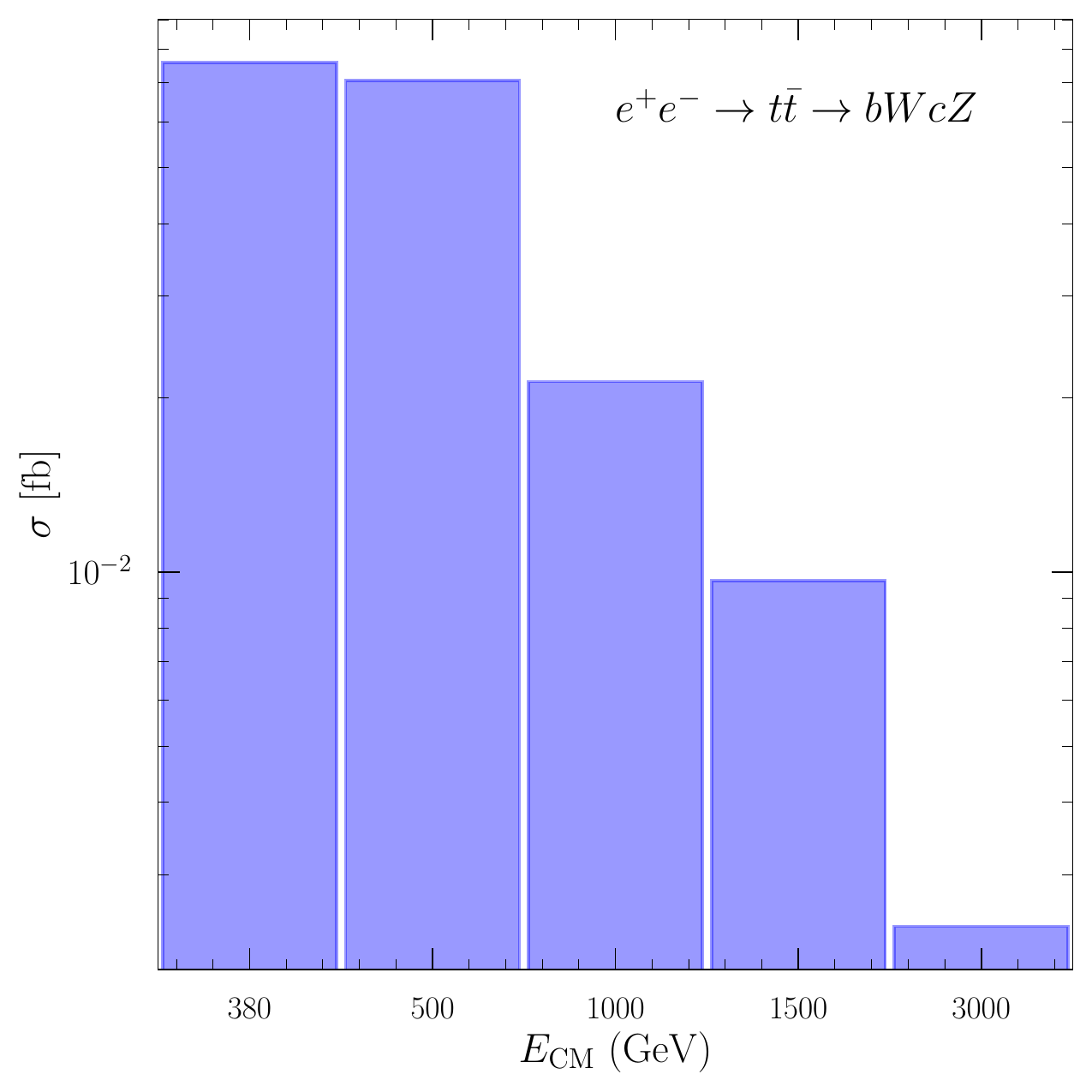}
    \caption{The total cross section for the production of $t\bar{t}$ followed by the decay into $bW cX$ as a function of the center-of-mass energy. We show the results for $cg$ (top left), $cH$ (top right), $c\gamma$ (bottom left) and $cZ$ (bottom right). All the calculations have been done at the lowest order in perturbation theory.}
    \label{fig:sigma:ee:tt}
\end{figure}

In Fig. \ref{fig:sigma:ee:tc}, we show the cross section for the production of a top (anti)quark in association with a charm (anti)quark for six center-of-mass energies $E_{\rm CM} = 250, 380, 500, 1000, 1500, 3000$ GeV which correspond to the expected runs for the International Linear Collider (ILC) and the Compact LInear Collider (CLIC). The production cross sections are estimated for the maximum allowed values of the effective couplings from the LHC searches and assuming a pure vectorial ($K_{qt}^L = K_{qt}^R = 0$) and tensorial ($X_{qt}^L = X_{qt}^R$) couplings. We can see that for the case of a pure vectorial $tqZ$ vertex, the cross section reaches a maximum of $0.76$ fb for $E_{\rm CM} = 380$ GeV. The $1/s$ behavior of the $tqZ$ contribution is not really dominant as the $tq\gamma$ contribution scales as $s$ due to the fact that the corresponding interaction is of a derivative nature. The case of a pure tensorial $tqZ$ vertex is more interesting as it leads to a cross section of order $3.5$ fb for $E_{\rm CM} = 500$ GeV and it slowly decreases with the center-of-mass energy. Note that with these numbers, we expect about $1400$-$7000$ FCNC events at Linear Colliders. With a small expected backgrounds and a clean environment, analyses of these phenomena at Linear Colliders can lead to impressive bounds. Note shown here, the cross section for the production of a top (anti)quark in association with an up (anti)quark is about a factor of two smaller since the LHC bounds on the corresponding $tuZ/tu\gamma$ are more stringent. 

The total rates for the production of $t\bar{t}$ where one top quark decays into $bW$ and the other one decays into $cX$ are shown in Fig. \ref{fig:sigma:ee:tt} for five representative center-of-mass energies starting from the $t\bar{t}$ threshold. Note that for this case, the production cross sections reach the maximum near the $t\bar{t}$ threshold with values of $0.5$ fb and $0.7$ fb for the $cg$ and $cH$ channels respectively. The cross sections for the $c\gamma$ and $cZ$ channels are about one order of magnitude smaller. With these numbers, we can get about ${\cal O}(1000)$ events at $E_{\rm CM} = 380$ GeV. The cross sections for the $uX$ channels are about a factor of two smaller.

\section{Conclusions}
\label{sec:conclusions}

In this work, we have discussed the top quark FCNC phenomena at future linear colliders. We first discussed a {\it novel} mechanism where the top quark FCNC decays are generated by the loops of dark sector particles. Within a simplified DM model which consists of two extra dark particles that are odd under an ad-hoc $Z_2$ symmetry: a colored scalar mediator carrying the same quantum numbers as a right-handed up-type quark and a right-handed fermion. The latter plays the role of the DM candidate. We have found that for phenomenologically viable scenarios, the model leads to sizeable rates for top quark FCNC decays into $qZ$ and $qH$. We then discussed the top quark FCNC phenomena in both the production ($t\bar{q}+{\rm h.c.}$) and in the decay ($t\bar{t} \to qXbW$) stages for the most generic dimension five FCNC interaction Lagrangian at future linear colliders. We found that for some processes, the production rates can be important which are about ${\cal O}(1)$ fb. Combining the relatively large production cross sections with the clean environment, the discovery prospects of top quark FCNC phenomena are promising but require more work in the future.

\section*{Acknowledgments}
The work of A.J. is supported by the Institute for Basic Science (IBS) under the project code, IBS-R018-D1. The work of S. K. was supported by the JSPS KAKENHI Grant No. 20H00160 and No. 23K17691.

\end{document}